# Formation of InAs Self-Assembled Quantum Rings on InP

T. Raz,[a] D. Ritter and G. Bahir

*Department of Electrical Engineering, Technion - Israel Institute of Technology,*

*Technion City, Haifa 32000, Israel*

## ABSTRACT

Shape transformations of partially capped self-assembled InAs quantum dots grown on InP are studied. Atomic force microscopy (AFM) images show large anisotropic redistribution of the island material after coverage by a 1 nm thick InP layer. The anisotropic material redistribution occurs within a few minutes and leads to a change from lens-like to elongated ring-like islands. The shape transformation is not accompanied by dot material compositional change. The formation of InAs/InP quantum rings disagrees with a previous model of InAs/GaAs ring formation that assumes that the driving force for the dot to ring transformation is the difference in surface diffusion velocity of indium and gallium atoms.

Electronic mail: talraz@tx.technion.ac.il

Atomic force microscopy and transmission electron microscopy reveal that the shape of conventional self-assembled semiconductor quantum dots (SAQDs) is lens-like in some cases,[1,2] and pyramid or truncated pyramid like, in other cases.[5,6] Capping of the dots by a layer thinner than the dot height, and a subsequent growth interruption, dramatically alters that shape of the SAQDs. Anisotropic redistribution of the quantum dot material takes place during the initial stages of the capping process and the following growth interruption. As a result, the quantum dots transform into elongated quantum rings with crater-like holes in their centers. The formation of ring shaped nano-structures was first reported for InAs partially capped islands (PCIs) on GaAs grown by molecular beam epitaxy (MBE),[7] and later for $In_{0.5}Ga_{0.5}As$ PCIs on GaAs grown by metalorganic vapor phase epitaxy (MOVPE).[8]

Since first observed, the ring shaped nano-structures were extensively studied experimentally[7-16] and theoretically.[17-21] The quantum rings are of much interest due to their electronic properties such as the large and negative excitonic permanent dipole moments,[9] the high oscillator strength of the band-to-band ground state transition,[10] and the possibility to tune their electronic states.[11] Initially, it was not clear whether the electronic structure inside the ring shaped nanostructures is indeed ring-like.[7,12] However, subsequent studies of their magnetic properties provided decisive evidence for the ring-like electronic structures.[13,14] The optical emission of a charged single quantum ring was also investigated.[15]

Two different models of the ring formation process have been suggested. The first model is based on kinetic considerations, namely on the differences in surface diffusion rate of indium and gallium atoms.[13,16] According to this model, mobile indium atoms diffuse from the dot outwards, leaving a void at the original dot location. The indium atoms diffuse into the surrounding GaAs crystal and form an immobile ring-shaped InGaAs island. The second model is based on thermodynamic considerations. It maintains that the change in surface free energy balance due to the thin layer overgrowth creates an outward pointing force.[16,22] This force brings about a material redistribution resulting in a ring shaped structure.

The similarity between semiconductor quantum ring structures and ring structures made of other materials such as liquid crystal and liquid metal films was already pointed out in Ref. 13. However, the In(Ga)As/GaAs system is the only semiconductor material systems where ring structures were reported. Though partially capped InAs islands on InP were studied, only height trimming effect of the quantum dots was observed, and no shape change to ring structure was reported.[23,24] This letter addresses the formation of InAs self-assembled quantum rings on InP. This is the first evidence that self-assembled quantum rings can be manufactured in a semiconductor material system other than In(Ga)As/GaAs. The observation of InAs self-assembled quantum rings on InP is inconsistent with the above-mentioned kinetic model of the dot to ring transformation.

The samples in this study were grown by a compact metalorganic molecular beam epitaxy (MOMBE) system[25] at 495ºC on semi-insulating InP(100). Trimethylindium, triethylgalium, arsine, and phosphine served as group III and V sources, respectively. A 200 nm thick InP buffer layer was first grown at a growth rate of 1.3 ML/s with a V/III ratio of 1.8. After a 40 sec growth interruption, the SAQDs were formed by depositing 2.1 ML of InAs at a growth rate of 0.11 ML/s with V/III ratio of 30. After another 40 sec growth interruption the SAQDs were partially capped by 1 nm thick InP layer at a growth rate of 0.5 ML/s with V/III ratio of 3.9. The samples were subsequently cooled, at a cooling rate of ~0.5°C/s, and removed from the system. Samples were scanned by Digital Instruments atomic force microscope (AFM) in the tapping mode. Samples for photoluminescence (PL) measurements were grown under the same conditions, with an additional InP cap layer over the partially capped dots. The growth was interrupted for 20 sec to 8 min in different samples, prior to the growth of the top InP cap layer. The InP cap layer was grown at the same conditions as the buffer layer, and was 50 nm thick. Photoluminescence (PL) spectra were measured at 77K using a HeNe laser for excitation, and a liquid nitrogen cooled Ge detector. As a reference sample of SAQDs were grown at the same conditions but without the 1 nm thick InP layer partially capping the dots.

The images obtained by AFM are shown in Fig. 1. The dot density in the uncapped reference sample (Fig. 1a) is about $3\times10^9$ cm$^{-2}$, and their shape is lens-like with an average base diameter of 75 nm and average height of 15 nm. The AFM image of dots that were capped by 1 nm InP (Fig. 1b) reveals ring-like structures elongated along [110] direction. The typical outer diameters of the rings are about 220 nm along the [110] direction, and about 110 nm along the [1$\bar{1}$0] direction. The average ring height is about 2.5 nm. Cross-section profiles taken along the [110] and [1$\bar{1}$0] directions are shown in Fig. 1c. The ring profiles were vertically shifted in the figure by 1 nm to take into account the thickness of the InP cap layer.

Table I compares the dimensions of the capped and uncapped dots reported here and the dimensions of InAs QDs grown on GaAs and capped by a 2 nm thick layer of GaAs reported elsewhere.[7] Both the capped and uncapped InAs dots are smaller when grown on GaAs. However, similar trends are observed in both material systems: the rings are elongated along the [110] direction, and the long diameter of the ring is about 3 times larger than the diameter of the uncapped lens-shaped dots, while the short diameter along the [1$\bar{1}$0] direction is about 1.5-2 times larger than the uncapped dot diameter. The average height is reduced by a factor of about 7 during the dot-to-ring transformation.

The PL spectra of the dots and rings (vertically displaced for clarity) are shown in Fig. 2. Both dots and rings were capped by 50 nm of InP in the samples that were grown for PL experiments. The lowermost curve (A) is the spectrum of the reference sample in which the 50 nm thick InP cap layer was grown immediately after the dots were formed. The other spectra (B, C, D) were obtained from samples with partially capped dots, with different growth interruptions between the partial capping by the 1 nm thick InP layer and the subsequent final capping by the 50 nm thick InP layer. The distinct PL peaks appearing at the high-energy side of the main peak in each spectrum are emitted from quantum dots whose height varies by a discrete number of monolayers.[26-28] The highest energy peak at 1.18 eV is emitted from a 2 monolayer high SAQDs, the next lower energy peak at 1.09 eV from a 3 monolayer high SAQDs, and so on. These distinct peaks are only slightly blue-shifted with increasing duration of the growth interruption, since the lateral dimensions of the dots or rings affect the peak position only marginally.

The dynamics of quantum ring formation is clearly manifested in the PL spectra. A short growth interruption of 20 sec after the dots are partially capped (spectrum B), results in no changes in the PL spectrum compared to that of the reference sample (A). Growth interruptions of 1 min (C) and 2 min (spectrum not shown here) cause merely a some broadening of the peaks emitted from 2 to 5 ML thick dots. However, after a 4 min long growth interruption the main peak shifts from 0.81 to 0.86 eV (D). This blue shift is the signature of the dot-to-ring transformation, when the high dots turn into much lower rings. No further changes in the spectra are observed after longer growth interruption of 6 or 8 min, indicating that the shape transformation is completed following a 4 min growth interruption.

In the previous reports on the formation of quantum rings in the InAs/GaAs material system it was found that material redistribution takes place within a few seconds to a few tens of seconds,[16] and therefore only a short growth interruption of less than 1 min is necessary to obtain quantum rings. By contrast, our PL spectra indicate that material redistribution associated with the dot-to-ring transformation in our experiments takes place on a time scale of a few minutes. This different behavior may be a result of the higher growth temperature of the samples studied in Ref. 16 (530ºC), in comparison to the growth temperature of the samples studied in this work (495ºC).

A major conclusion that can be drawn from our results is that the kinetic model which assumes that the difference in surface diffusion rate of Ga and In atoms leads to the transformation of InAs dots to InGaAs rings[13,16] does not apply here. In the InAs/InP material system the dot and the surrounding material share the same group III atom. Moreover, since the distinct peaks in the PL spectra exhibit only a slight blue shift during the growth interruptions (see Fig. 2), no compositional changes during the material redistribution take place. We therefore believe that the recently proposed thermodynamic model provides a better explanation for the formation of the rings. This model is based upon the difference of the surface free energy balance between the dot, the surrounding material, and the chamber atmosphere, while the dots are partially capped.[16,22] It predicts a dot-to-ring shape change not necessarily accompanied by a compositional change.

In conclusion, partially capped InAs SAQDs grown on InP exhibit a shape transformation from lens-shaped to ring-shaped islands. This shape transformation takes place on a time scale of a few minutes at a temperature of 495ºC. The ring-shaped islands formation results from a redistribution of the InAs only, with no compositional changes. The shape transformation is best explained by the thermodynamic model, which asserts that the ring-shaped islands are the equilibrium configuration due to the changes in surface free energy balance during the thin layer overgrowth.

**Figure Captions:**

Fig. 1: AFM images of InAs (a) uncapped and (b) partially capped dots grown on InP. (c) Cross-section profiles of the capped and uncapped dots along the $[110]$ and $[1\bar{1}0]$ directions.

Fig. 2: PL spectra obtained from 50 nm InP capped samples, following a growth interruption of (A) 0 sec, (B) 20 sec, (C) 1 min and (D) 4 min after the dots are partially capped.

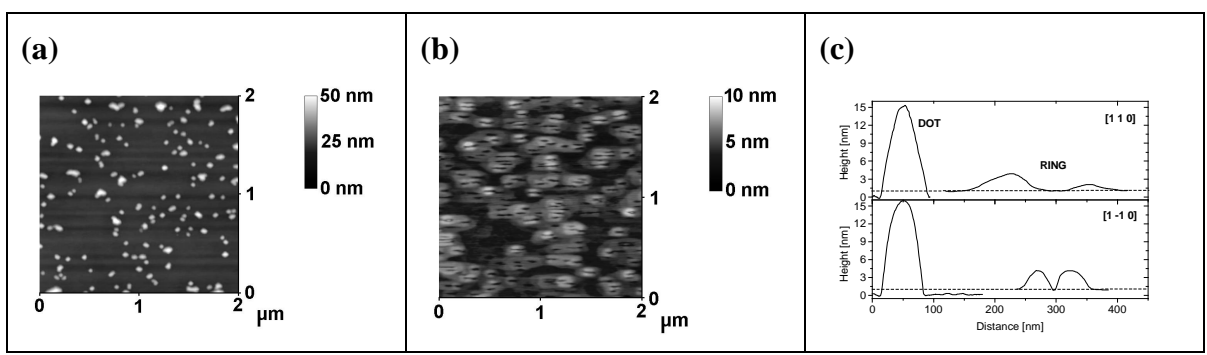

**Fig. 1**

T. Raz et al.

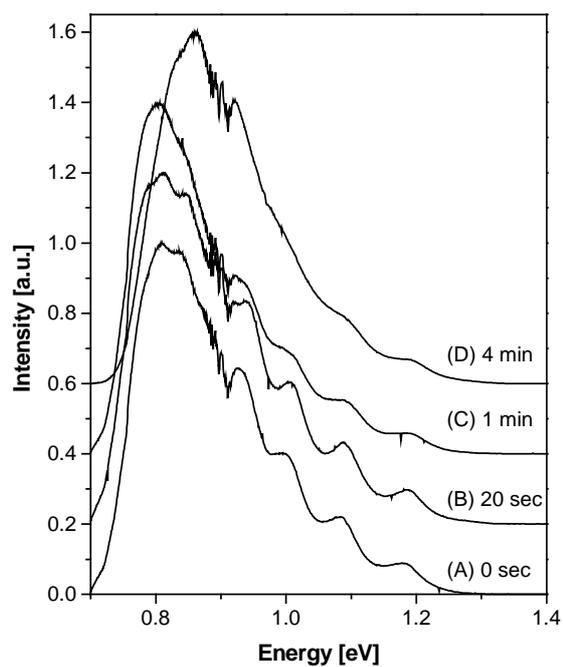

**Fig. 2**

T. Raz et al.

Table I. Dimensions [nm] of the uncapped and partially capped islands in the InAs/InP (this study) and InAs/GaAs[7] material systems. Values from Ref. 7 are for 2 nm caps, which is the smallest reported thickness.

|  |  | InAs/InP | InAs/GaAs[7] |
|---|---|---|---|
|  | Height | 15 | 10 |
| Uncapped islands | Length [110] | 75 | 34 |
|  | Length [1$\bar{1}$0] | 75 | 34 |
|  | Height | 2 | 1.5 |
| Partially capped islands | Length [110] | 220 | 110 |
|  | Length [1$\bar{1}$0] | 110 | 70 |